\documentstyle[11pt,newpasp,twoside,epsf]{article}

\begin{document}

\title{Observed Properties of Mass Loss in Symbiotic Binaries}

\author{Joanna Miko{\l}ajewska}

\affil{Copernicus Astronomical Center, Bartycka 18, PL-00716 Warsaw,
Poland, e-mail: mikolaj@camk.edu.pl}

\begin{abstract}
Both the red giants and the Mira variables
in symbiotic systems have systematically higher mass-loss rates
than do typical galactic giants and Miras,
which suggests that only very evolved giants, and so those with
highest mass-loss rates, can support symbiotic behaviour
in widely separated binary systems.
They often show a flattened mass-loss geometry
due to an intrinsically inhomogeneous 
mass loss and/or tidal interactions between the binary components.
The main body of a symbiotic nebula is thus formed from material
lost in the giant wind, while the hot component is responsible
for its ionization and excitation. In addition, the fast wind and/or
jet ejection from
the hot component, whenever occur, 
give rise to the complex, often bipolar, shape of 
symbiotic nebulae.
Observations of resolved nebulae also suggest that 
the binary geometry and nebular structure are aligned
but the bipolar outflow may be not orthogonal to 
the orbital plane in all cases.
\end{abstract}

\keywords{symbiotic binaries, circumstellar matter, mass loss}

\section{Introduction}

Symbiotic stars are interacting binaries composed of
an evolved giant primary and a hot, luminous companion
surrounded by an ionized nebula. 
Depending on the nature of the giant 
we have two distinct classes:
the S-type with normal giants and orbital periods
of about 1--15 yr, and the D-type with Mira primaries
usually surrounded by a warm dust shell, and orbital periods
generally longer than 10 yr.
The hot star in most cases appears to
be a white dwarf powered by thermonuclear
burning of the material accreted from its companion's wind.
The presence of both the evolved giant, heavily losing mass
in most cases, and the hot companion copious in ionizing photons
and often possessing its own wind lends large variety to the
circumstellar environment of symbiotic binaries.

This paper gives a brief summary of the observed properties
of the symbiotic
circumstellar envelopes. 
More details
can be found in Miko{\l}ajewska (1999, hereafter M99),
while for the most recent general review of the symbiotic stars
the interested reader is referred to Miko{\l}ajewska (1997).

\section{A simple case: cool giant wind ionized by the hot
companion}

So far $\sim 25 \%$ of all symbiotics have been detected
at radio range. In practically all cases, the radio emission 
at cm wavelengths is consistent with optically thick
thermal bremsstrahlung from an ionized gas (Seaquist \& Taylor 1990).
According to a simple binary model, 
the radio emission
originates from the red giant wind
ionized by the hot companion
(Seaquist et al. 1984). The geometry of this emission region
is governed by a single parameter, $X$, which depends
on the red giant mass-loss rate, the binary separation,
and the Lyman continuum luminosity of the hot component.
According to the model, the optically thick spectral index
depends on $X$, while 
the turnover frequency, $\nu_{\rm t}$, is related
to the binary separation. 
The model can be thus easily tested by observing 
of the radio spectra, determining $\nu_{\rm t}$, and then comparing 
the model-dependent estimate of the binary separation with independently
known values.

Preliminary results of such tests based on mm/sub-mm observations
of $\sim 40$ symbiotics
show that the model works
for quiet, non-variable S-type systems, and eruptive, S-type systems
during quiescence (M99; Miko{\l}ajewska
\& Ivison 1999, hereafter MI99). The only exception is CI Cyg,
one of the few symbiotic systems in which the M giant shows strong
tidal distortion, and loses mass via Roche-lobe overflow
rather than via stellar wind. 
Most of these systems seem to have 
$ 0.2 \la X < \pi/4$, which is consistent
with cone shaped ionization 
front with opening angle increasing with $X$.
Similar geometry is implied by studies of Raman scattered O\,{\sc vi} 
$\lambda\lambda$ 6825, 7082 emission lines observed
in many symbiotic stars (Schmid 1996).
The radio observations suggest
also that symbiotic giants have systematically higher mass-loss
rates than do normal red giants.
The hot component luminosity, $L_{\rm h}$, is roughly
correlated with the giant mass-loss rate, which
has a natural explanation in the frame
of proposed models for the hot component. 
Most symbiotics interact by wind accretion, and in
general, the stronger is the cool giant wind the more material
can be accreted  by its white dwarf companion.
As the expected accretion rate is a few per cent of $\dot{M}_{\rm wind}$,
only systems with low $L_{\rm h} < 100 \rm L_{\odot}$ (e.g. EG And) 
can be powered solely by accretion. To power a typical
$L_{\rm h} \sim 1000 \rm L_{\odot}$, the symbiotic white dwarfs must 
burn H-rich
material as they accrete it.
The latter is possible only if the accretion rate exceeds some
minimum value.
In both cases, the resulting $L_{\rm h}$ is related in someway to
the giant mass-loss rate.
The mass-loss rates derived for the symbiotic giants 
are sufficient to power the observed 
luminosities of their hot companions via wind-accretion.

\section{Dust envelopes of symbiotic Miras}

The near-IR colours of D-type symbiotic systems indicate
presence of warm dust shells, with $T_{\rm d} \sim 1000$ K
(e.g. Whitelock 1987).
In at least 50\% of studied systems, extinction for 
the Mira component is grossly different 
from that for
high excitation regions (emission lines, hot UV continuum),
and in {\it all} of them, E$_{B-V}({\rm Mira}) > {\rm E}_{B-V}(\rm hot)$
(e.g. M99).
The hot component in these systems must therefore lie
outside the dust cocoon of the Mira.
Assuming a typical dust formation radius of $\ga 5 R_{\rm c}$,
and the Mira radius, $R_{\rm c} \sim$ 2--3 a.u., this
immediately implies minimum binary separations $\ga$ 10--15 a.u., and periods
$\ga 20$ yr.

The IR light curves of well-studied D-type systems
show, in addition to periodic pulsation
of the Mira, significant long term variations
of the mean light level (Whitelock 1987; M99). 
RX Pup, which light curve has been recently analyzed by Miko{\l}ajewska
et al. (1999a),
is typical of these. 
The changes have decreasing amplitude with increasing wavelength
suggesting obscuration by dust. The Mira amplitude is practically
unaffected indicating the changes are due only to increased
extinction. In RX Pup, V835 Cen and He2-38, 
the changes in the reddening towards
the Mira are {\it not} correlated with similar changes
in the reddening towards the hot component and emission
line formation region(s).
The obscuration is also not accompanied by any
related changes in temperature or luminosity of the hot
component, and it apparently affects only the Mira.
The physical nature of this phenomenon remains a mystery.
In the specific case of R Aqr the obscuration was explained
in terms of orbitally related eclipses of the Mira
by pre-existing dust, and it has been suggested that
the events in other symbiotic Miras are similarly caused
(Whitelock 1987).
There is, however, ample observational evidence that the dust 
obscuration phenomenon in symbiotic Miras cannot
be, in general, orbitally related (M99,
and references therein).
In particular, 
the occurrence of similar changes in single Miras 
points to intrinsic variations in the Mira envelope.

The occurrence of dust obscuration phases may be related to 
intensive mass loss. In carbon Miras, the phenomenon seems
to favour objects with moderate dust shells.
The same seems to hold for symbiotic Miras.
In particular, known periods for symbiotic Miras range from 
280 to 578 day with a mean value
of about 425 day (Whitelock 1987), much
larger than the median value of
250-300 day estimated for single galactic Miras.
They have also redder $K$-[12] colours than average galactic Miras.
This suggests that only long-period Miras, and so those with highest
mass-loss rates, support symbiotic behaviour in widely separated
binary systems.
Symbiotic Miras with erratic IR variability have all 
colours between 3 and 5, and consequently
$\dot{M}_{\rm c} \sim {\rm a\,few}\,\times\,10^{-6}\, \rm M_{\odot}/yr$,
significantly higher than 
a characteristic rate of $\sim 10^{-7}\, \rm M_{\odot}/yr$
derived for O-rich single Miras with periods in the range of 
200-400 day (Jura \& Kleinman 1992).

In RX Pup and other symbiotic Miras,
there is also no evidence for reprocessing of the Mira light
during the obscuration phase since weakening at shorter wavelengths
($J,H$) is not accompanied by brightening at longer
wavelengths ($L$). The same is observed in R For
and other C Miras, and can be hardly reconciled with a spherically 
symmetric dust ejection (Whitelock et al. 1997).
The alternative scenarios involve ejection around equatorial
disk or as puffs in random direction.

\section{Hot component wind and geometry of resolved symbiotic nebulae}

The hot components very often show evidence
for some mass-loss associated with their activity, such as thermonuclearly
or accretion-powered eruptions, or unstable accretion onto magnetic
white dwarf.
In particular, all recorded symbiotic novae eruptions were followed 
by a phase of intensive Wolf-Rayet-type wind from the hot component.
Ejection of jets from the hot component of R  Aqr 
and CH Cyg has been recorded in the radio and optical.
There is also evidence for the existence of such winds
in quiescent systems.
The wind velocity is usually in the range $\ga$ 200--1000 km\,s$^{-1}$.
There have been only a few attempts to estimate the mass loss
rates. For the symbiotic novae, 
these estimates range from 
$\dot{M}_{\rm h} \sim 10^{-5}\, \rm M_{\odot}/yr$ in RX Pup 
(Miko{\l}ajewska et al. 1999a) and $\la 10^{-5}\, \rm M_{\odot}/yr$
in PU Vul (Sion et al. 1993) to 
$3\,\times\,10^{-7}$--$10^{-6}\, \rm M_{\odot}/yr$ in AG Peg
(Vogel \& Nussbaumer 1994; Kenyon et al. 1993),
and they roughly scale with their luminosity during the plateau phase.
Both in AG Peg and RX Pup the intensity of the wind diminished
in step with the hot component lumininosity 
during the decline of the outburst.
Similarly, the estimates of $\dot{M}_{\rm h} \sim 10^{-8}\, \rm M_{\odot}/yr$
in BF Cyg (Miko{\l}ajewska et al. 1989),
$\sim 5\,\times\,10^{-9}\, \rm M_{\odot}/yr$ in
EG And (Vogel 1993) and 
$\sim 10^{-7}\, \rm M_{\odot}/yr$ 
in AE Ara (Miko{\l}ajewska et al. 1999b) show some correlation with
the hot component luminosity.

Whenever the wind from the hot component occurs it 
should collide with the cool giant wind, giving rise
to high energy phenomena. In fact, 
among systems mentioned in this section,
EG And, RX Pup, CH Cyg, PU Vul, AG Peg and R Aqr have been detected
by ROSAT, and all but R Aqr show the $\beta$-type spectra
that can be reproduced with the emission
from a very hot, $T \ga 10^6$ K, optically thin plasma
possibly heated by the shocks in the collision of two stellar
winds (M{\"u}rset et al. 1997). 
These findings are very important as the interaction of winds from
the two binary components can play a major role in the final
appearance of symbiotic nebulae. 
According to the most popular scenario, a bipolar nebula is formed
under the action of a fast wind from the central object 
expanding in asymmetric AGB remnants.
The symbiotic interacting binary system thus offer the most
natural environment for production of such a bipolar nebula.
The properties of dust envelopes of symbiotic Miras 
discussed above are consistent with the required flattened mass-loss
geometry, while the presence of the fast wind from the hot 
component, at least occasionally, seems to be a common property
of symbiotic systems. 

Unfortunately, due to small relative
sizes, only 17 symbiotic nebulae
are thus far resolved in the optical and/or radio 
(e.g. Corradi et al. 1999),
and only in the case of R Aqr the binary itself 
has been spatially resolved. 
Although only about 20\% of all known symbiotic binaries 
are D-type systems with a Mira primary, most of extended,
radio and/or optically resolved ionized nebulae are associated with 
symbiotic Miras and active S-type systems (Corradi et al. 1999).
A picture gallery of these nebulae presented by Corradi \& Schwarz
(see Miko{\l}ajewska 1997)
reveals their generally complex structure, often with bipolar
lobes and jet-like components. 
Five of these complex nebulae, RX Pup, HM Sge, V1016 Cyg, V1329 Cyg
and AG Peg, are associated with recorded symbiotic nova eruption,
and two of them, CH Cyg and R Aqr, with ejection of jets.

For some systems with resolved nebulae, the orientation of the binary
on the sky can be derived from polarimetric studies,
allowing to distinguish between polar and equatorial outflows.
The formation scenarios can be thus critically tested and revised.
The scattering geometry and nebular structure are aligned in all
cases although the bipolar outflow in some cases
may be not perpendicular to the orbital plane
(M99, and references therein).

\acknowledgments
I would like to thank the LOC for their support.
This work was partly supported by Polish KBN Research Grant 
No. 2\,P03D\,021\,12.


\begin{references}

\reference Corradi, R.L.M., Brandi, E., Ferrer, O.E., Schwarz, H.E.,
           1999, \aap, 343, 84
\reference Jura, M., Kleinman, S.G., 1992, \apjs, 79, 105
\reference Kenyon, S.J., Miko{\l}ajewska, J., Miko{\l}ajewski, M.,
           et al., 1993, \aj, 103, 1573	   
\reference Miko{\l}ajewska, J. (ed.), 1997, Physical Processes 
           in Symbiotic Binaries and Related Systems, 
           Copernicus Foundation for Polish Astronomy, Warsaw
\reference Miko{\l}ajewska, J., 1999, in Stecklum B. et al. (eds),
           Optical and Infrared Spectroscopy of Circumstellar Matter,
	   ASP Conf. Series, in press (M99)
\reference Miko{\l}ajewska, J., Ivison, R.I, 1999, in Charles et al. (eds.),
           Warner Symposium Proceedings, in press (MI99). 
\reference Miko{\l}ajewska, J., Kenyon, S.J., Miko{\l}ajewski, M., 1989,
           \aj, 98, 1427	   
\reference Miko{\l}ajewska, J., Brandi, E., Hack, W., et al., 
           1999a, \mnras, 305, 190
\reference Miko{\l}ajewska, J., Belczy{\'n}ski, K., 
           Brandi, E., et al., 1999b, \aap, submitted
\reference M{\"u}rset, U., Wolff, B., Jordan, S., 1997, \aap, 319, 201
\reference Nussbaumer, H., Vogel, M., 1994, \aap, 284, 145
\reference Sion, E.M., Shore, S.N., Ready, C.J., Scheible, M.P., 1993,
           \aj, 106, 2118
\reference Schmid, H.M., 1996, \mnras, 282, 511
\reference Seaquist, E.R., Taylor, A.R., 1990, \apj, 349, 313 
\reference Seaquist, E.R., Taylor, A.R., Button, S.,1984, \apj, 284, 202
\reference Vogel, M., 1993, \aap, 274, L21
\reference Whitelock, P.A., 1987, \pasp, 99, 573 
\reference Whitelock, P.A., Feast, M.W., Marang, et al., 1997,
           \mnras, 288, 512
\end{references}
\end{document}